\documentclass[aps,prl,twocolumn,showpacs,showkeys,groupedaddress]{revtex4-1}
\usepackage{multirow}
\usepackage{graphics,graphicx}

\begin{document}

%\preprint{}

%\title{The nature of impurity-mediated grain boundary deformation in $fcc$ copper}
%\title{Impurity effects on the grain boundary deformation behavior in copper}
%\title{\textit{Ab initio} theory of impurity-induced ductilization or embrittlement of polycrystalline copper}
%\title{The mechanism of impurity influence on the ductility of grain boundaries in copper}
\title{Impurity effects on the grain boundary cohesion in copper}
%\title{Predicting the effect of impurity segregation on the mechanical behavior of grain boundaries in copper}


\author{Yunguo Li}
\email[Email (Yunguo Li): ]{yunguo.li@ucl.ac.uk}
%\thanks{}
\affiliation{Department of Materials Science and Engineering,\\ Royal Institute of Technology (KTH), S-100 44 Stockholm, Sweden}
\altaffiliation[Present address: ]{Faculty of Mathematical and Physical Sciences, University College London, Gower Street, London WC1E 6BT, United Kingdom}

\author{Pavel A. Korzhavyi}
\email[Email (Pavel A. Korzhavyi): ]{pavelk@kth.se}
\affiliation{Department of Materials Science and Engineering,\\ Royal Institute of Technology (KTH), S-100 44 Stockholm, Sweden}
\affiliation{Institute of Metal Physics, Ural Division of the Russian Academy of Sciences, 620219 Ekaterinburg, Russia}

\author{Rolf Sandstr{\" o}m}
\affiliation{Department of Materials Science and Engineering,\\ Royal Institute of Technology (KTH), S-100 44 Stockholm, Sweden}

\author{Christina Lilja}
\affiliation{Swedish Nuclear Fuel and Waste Management Company, Box 250, SE-101 24 Stockholm, Sweden}


\date{\today}

\begin{abstract}
Segregated impurities at grain boundaries can dramatically change the mechanical behavior of metals, while the mechanism is still obscure in some cases. Here, we suggest an unified approach to investigate segregation and its effects on the mechanical properties of polycrystalline alloys using the example of 3$sp$ impurities (Mg, Al, Si, P, or S) at a special type $\Sigma 5(310)[001]$ tilt grain boundary in Cu. We show that for these impurities segregating to the grain boundary the strain contribution to the work of grain boundary decohesion is small and that the chemical contribution correlates with the electronegativity difference between Cu and the impurity. The strain contribution to the work of dislocation emission is calculated to be negative, while the chemical contribution to be always positive. Both the strain and chemical contributions to the work of dislocation emission generally become weaker with the increasing electronegativity from Mg to S. 
%We find S to be an embrittler while P not to embrittle Cu, which matches well with experiments. The qualitative contrast between the effects of P and S on the ductility of grain boundaries in Cu is found to be a result of accumulated quantitative differences of the basic physical contributions.
By combining these contributions together we find, in agreement with experimental observations, that a strong segregation of S can reduce the work of grain boundary separation below the work of dislocation emission, thus embrittling Cu, while such an embrittlement cannot be produced by a P segregation because it lowers the energy barrier for dislocation emission relatively more than for work separation.

\end{abstract}

\pacs{62.20.Mk, 61.72.Mm, 61.72.Bb, 68.35.Dv, 05.70.Np}
%\keywords{}

\maketitle

Impurity-induced embrittlement accounts for many notorious cases of brittle failure of polycrystalline metals \cite{duscher-Bi-Cu,Ludwig2005151,laporte2009intermediate}. Bismuth-embrittled nickel and copper are well-known cases of such an embrittlement behavior and, therefore, they have been extensively studied \cite{duscher-Bi-Cu,schweinfest2004bismuth,lozovoi2006structural,PhysRevLett.111.055502}. The impurity-induced embrittlement was attributed either to a chemical effect of the Bi segregation, which is believed to change the bonding strength at grain boundaries (GBs) \cite{duscher-Bi-Cu,PhysRevLett.111.055502}, or to a size (strain) effect that is associated with the size misfit of Bi in the Cu lattice \cite{schweinfest2004bismuth,lozovoi2006structural}. However, no theory could explain the remarkable difference between the effects of P and S on the ductility of polycrystalline copper \cite{henderson92tr,sandstrom2013influence}. Segregated S at GBs is strongly detrimental and several ppm of residual S can remarkably embrittle copper \cite{suzuki1984intermediate,fujiwara1995ductility,laporte2009intermediate}. However, the addition of about 50 wt. ppm of the neighboring element P can cure the Cu embrittlement problem and recover the ductility of polycrystalline copper \cite{sandstrom2013influence}. Evidently, the atomistic mechanism of grain boundary deformation with segregated impurities needs further clarification.

In this letter, we suggest a unified approach based on first-principles calculations with which the segregation of 3$sp$ impurities at extended defects and the segregation effects on mechanical behavior of polycrystalline copper can be investigated. Our analysis shows that Mg, Al and Si do not embrittle Cu, P can improve the ductility of polycrystalline copper, and S can cause intergranular embrittlement of Cu. The chemical and size effects on both the work of GB decohesion (also called work of separation $W_{sep}$) and the work required for dislocation emission (dislocation nucleation threshold $G_{disl}$) generally follow the change of electronegativity ($\chi$) for the 3$sp$ impurities in Cu. The sharp contrast between the effects of P and S on the mechanical behavior of GBs in Cu is found to result from the quantitative differences between the effects of these impurities on the $W_{sep}$ and $G_{disl}$. These findings explain well the experimental observations and may stimulate the efforts aimed at grain boundary engineering in polycrystalline metals.

Nucleation, growth, and coalescence of cracks or voids at GBs are the basic processes involved in the creep deformation and intergranular failure of polycrystalline metals \cite{THOMSON19861}. The deformation behavior of GBs may be brittle or ductile depending on how cracks propagate along them, which in turn is a result of the competition between the events of crack advance (by brittle cleavage) and crack blunting (by dislocation emission) associated with an atomically sharp crack tip \cite{cheng2010intrinsic,armstrong1966cleavage,kelly1967ductile,rice1974ductile,mason1979segregation,rice1992dislocation,phillpot1999synthesis,yamakov2006molecular}. This competition is well illustrated in the model by Rice, Thomson, and Wang \cite{rice1992dislocation,rice1974ductile,rice1989embrittlement}, which states that a GB crack propagates in a brittle or a ductile fashion depending on whether the specific energy release rate required to emit a single dislocation ($G_{disl}$), is higher or lower than that associated with the brittle GB decohesion ($W_{sep}$).  

A polycrystal can lower its energy by accumulating impurities at stacking faults (SFs), dislocations and GBs. As a consequence, impurities may form so rich segregations at extended defects that these regions become qualitatively different from the host crystal in terms of chemical bonding, which may induce dramatic changes of material's properties. To investigate how impurities change the GB deformation behavior, we must calculate both $W_{sep}$ and $G_{disl}$ in the absence or presence of impurities at GBs. The ideal work of separation is given by $W_{sep} = 2\gamma_s - \sigma$ according to Griffith's fracture theory \cite{griffith1921phenomena}, where $\sigma$ and $\gamma_s$ are the surface energies for the GB and for each of the two opened surfaces.
According to Rice \cite{rice1992dislocation}, the threshold for dislocation nucleation at a GB crack under tensile loading normal to the GB/crack plane (mode I) may be related to the unstable stacking fault (USF) energy $\gamma_{usf}$ as
\begin{eqnarray}\label{eq:ricemd} 
G_{disl} = 8\gamma_{usf}[1 + (1-\nu)\rm{tan}^2\phi] / [(1+ \rm{cos}\theta)\rm{sin}^2\theta],
\end{eqnarray}
where $\nu$ is the Poisson's ratio, $\theta$ is the inclination angle of the slip plane with respect to the crack plane, and $\phi$ is the angle between the Burgers vector and the normal to the crack front in the slip plane. The surface energies $\sigma$, $\gamma_s$ and $\gamma_{usf}$ can be calculated as derivatives $\left( \frac{\partial G}{\partial A} \right)_{T, P, n_i} $ of the respective excess Gibbs free energy $G$ with respect to the surface area $A$. Here $T$ and $P$ are the temperature and pressure, respectively, and $n_i$ is the number of atoms of species $i$ in the system. In the presence of impurities, an additional contribution to in the Gibbs free energy, $\Delta G_{seg}$ due to the impurity segregation, will appear in the numerator.

Let us consider the case of strong segregants at low temperature $T \to 0$ to neglect the entropy term and also take $P=0$. Then the the Gibbs free energy may be accurately approximated by the total energy $E$ of a static atomic configuration, which can be calculated using density functional theory (DFT) in the generalized gradient approximation (GGA-PBE) \cite{perdew1996generalized} as implemented in the Vienna Ab-initio Simulation Package (VASP) \cite{kresse1999ultrasoft, blochl1994projector}, see Supplemental Material.

\begin{figure}[htb!]
\centering
\includegraphics[width=0.5\textwidth]{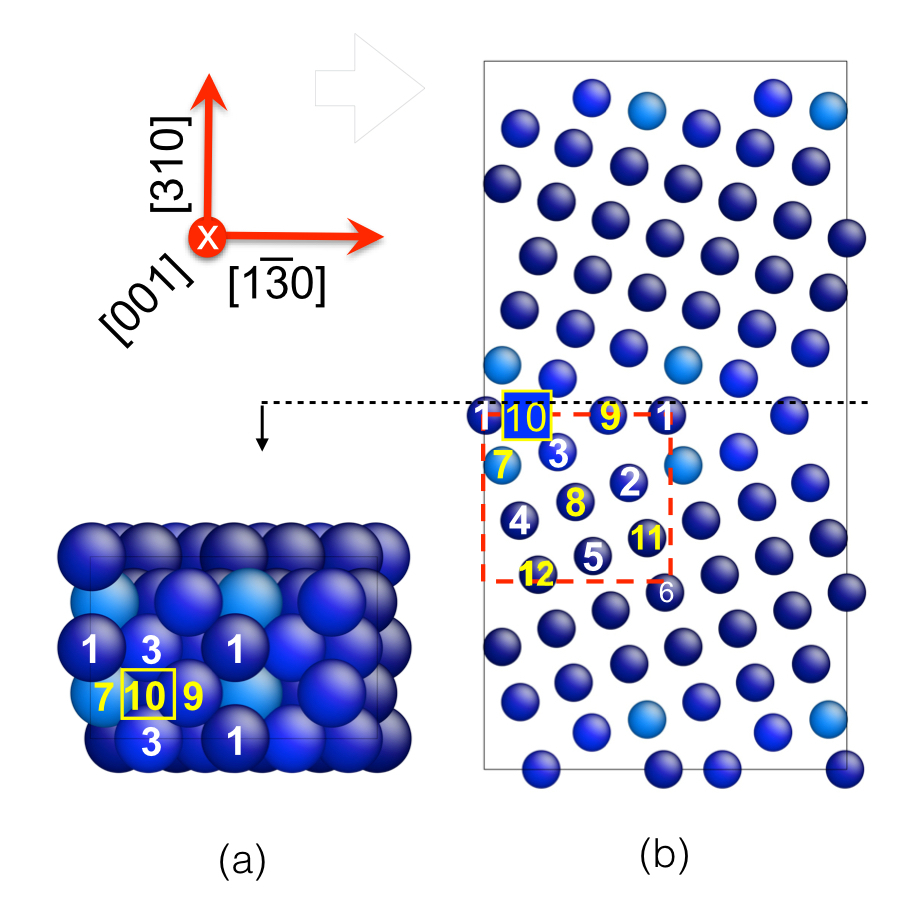}
\caption{Atomic structure of the $\Sigma5(310)[001]$ symmetric tilt GB in a supercell model comprised of 152 atoms: top view (a) and side view (b). The inequivalent sites are labeled by numbers. Atoms are colored according to the magnitude of atomic local strain tensor  \cite{shimizu2007theory}, with dark blue indicating zero and light blue indicating higher value. \label{fig:gbstr}}
\end{figure}

We model the atomic structure of $\Sigma 5(310)[001]$ symmetric tilt GB by a periodic supercell with two identical grain boundaries separated by nineteen (310) layers as shown in Fig. \ref{fig:gbstr}: This GB is chosen for the present study because it has a relatively high energy (0.871 J/m$^2$) similar to the energy of random GBs and is abundant in polycrystalline copper. We simulate  isolated impurities using a 3$\times$3$\times$3 fcc Cu-based supercell (108-atoms) with one substitutional impurity atom inside. The calculated $\sigma$, $\gamma_s$, $\gamma_{usf}$, $W_{sep}$, and $\Delta G_{seg}$ are in good agreement with experimental data and results of previous calculations. Details of the present GB, SFs and free surface modeling can be found in the Supplemental Material. 
\begin{table}[htb]
\centering
 \caption{Segregation energies of the 3$sp$ elements to the $\Sigma5(310)[001]$ GB, (310) open surface, ISF, and USF of copper, calculated using PBE functional. The negative sign indicates that the segregation site is energetically preferred to the bulk.  \label{tbl:seg}}
\begin{tabular} {@{\extracolsep{6pt}}lccccc@{}}
\hline
  \multirow{2}{*}{Site} & \multicolumn{5}{c}{Segregation Energy (eV/atom)} \\
\cline{2-6}
%& \multicolumn{5}{c}{\textit{fixed-composition}} &\multicolumn{5}{c}{\textit{variable-composition}} \\ 
& Mg&Al &Si &P &S  \\
\cline{1-6}
 1 &-0.25 &-0.29 &-0.58 &-0.90 &-1.03 \\
 2 &-0.42 &-0.29 &-0.29 &-0.43 &-0.58 \\
 3 &-0.32 &-0.30 &-0.57 &-0.95 &-1.15 \\
 7 &-0.62 &-0.28 &-0.32 &-0.35 &-0.69 \\
 9 &-1.13 &-0.47 &-0.05 &-0.25 &-0.68 \\
 10 &0.33 &0.01 &-0.54 &-0.93 &-0.76 \\
(310)$_s$  &-0.63 &0.10 &-0.22 &-1.14 &-2.01\\
ISF &-0.047 &-0.053 &-0.100 &-0.131 &-0.136\\
USF &-0.030 &0.007 &-0.026 &-0.092 &-0.038\\
\hline   
\end{tabular}
\end{table}
\begin{figure}[h!]
\centering
\includegraphics[width=0.48\textwidth]{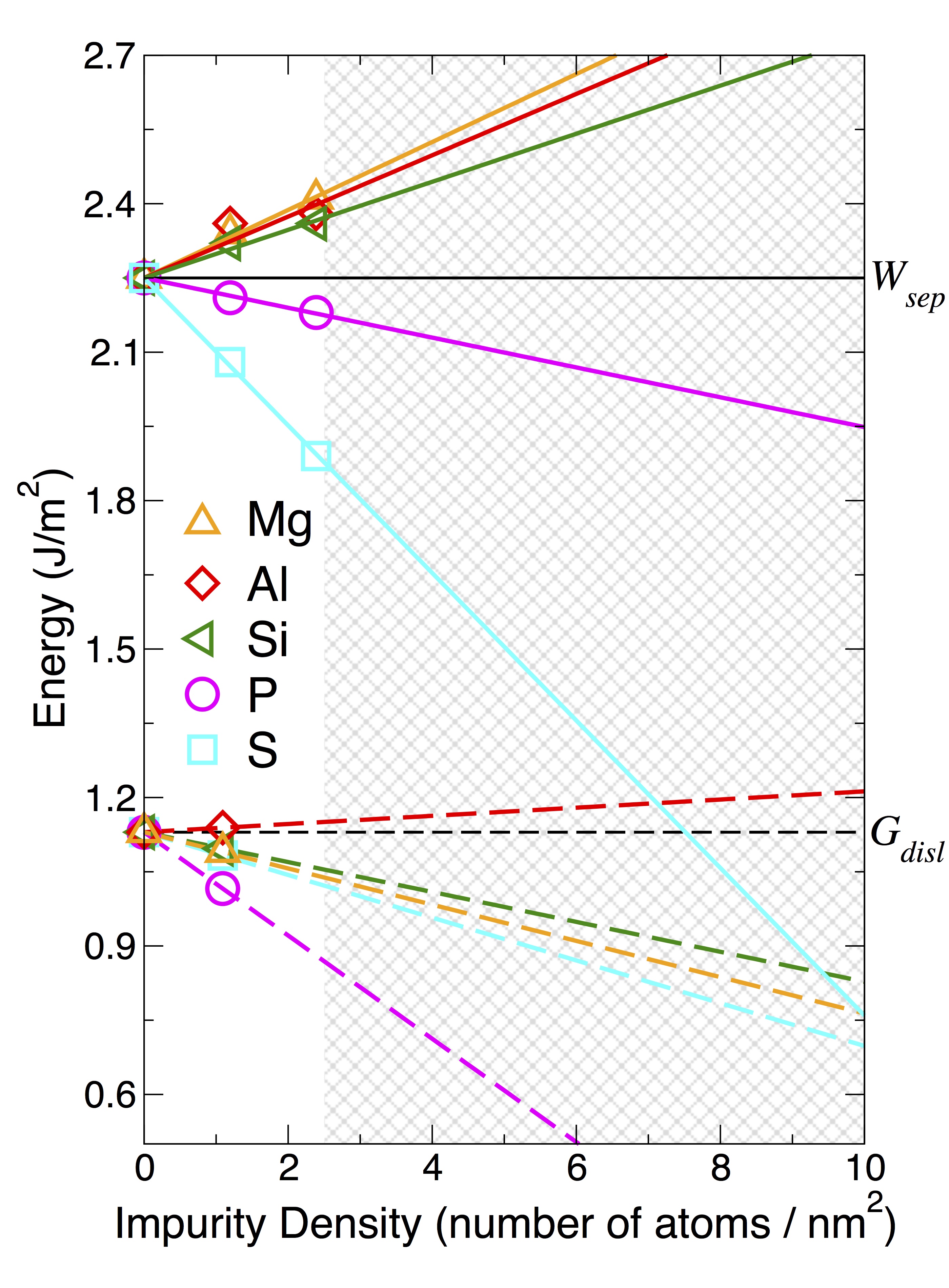}
\caption{Segregation-induced changes of the work of grain boundary separation and the work of dislocation nucleation as a function of impurity segregation density. Symbols are the calculated data. Lines are obtained by linear fit. The cross-hatching indicates the area to which the results have been extrapolated. \label{fig:energ-den}}
\end{figure}
\begin{table}[h!]
\centering
 \caption{Segregation-induced changes of $W_{sep}$ and $G_{disl}$ with contributions from the chemical (\textit{chem}) and relaxation (\textit{strain}) effects. \label{tbl:wkg-dil}}
\begin{tabular} {@{\extracolsep{6pt}}lccccccc@{}}
\hline
 \multirow{2}{*}{Impurity} & \multicolumn{3}{c}{$\Delta W_{sep}$ (mJ/m$^2$)} & \multicolumn{3}{c}{$\Delta G_{disl}$ (mJ/m$^2$)} \\
\cline{2-7}
   & \textit{total} & \textit{chem}  & \textit{strain} & \textit{total} & \textit{chem}  & \textit{strain}\\
\hline
Mg &94.0 &81.2  &12.8 &-37.7 &419.8 &-457.5\\
Al &110.0 &112.0  &-2.1 &8.8 &328.0 &-319.2 \\
Si &68.1 &75.1  &-7.0 &-33.0 &182.7 &-215.7\\
P &-36.9 &-35.3  &-1.5 &-114.6 &92.9 &-207.5 \\
S &-165.4 &-178.4  &12.9 &-47.1 &101.7 &-148.8 \\
\hline   
\end{tabular}
\end{table}

To evaluate the impurity segregation effects on $W_{sep}$ and $G_{disl}$ we calculated the impurity segregation energies to a $\Sigma5(310)[001]$ symmetric tilt GB, a (310) surface, and two types of stacking faults (SFs) in copper using supercell models of these defects, first in pure copper and then with one or two impurity atoms in the defected region (for details, see Supplemental Material). The obtained results were then extrapolated to higher impurity concentrations. In the calculations for the SFs almost linear concentration dependencies were obtained implying that the interactions among the impurities mostly cancel out. We therefore assume a linear dependence of $G_{disl}$ on the impurity segregation density parameter $\Gamma$.

Table \ref{tbl:seg} reports the calculated segregation energies for the 3$sp$ impurities towards a $\Sigma5$ GB, a (310) open surface, an intrinsic SF, and an unstable SF. All the impurities are found to prefer to segregate to the GB rather to the SFs. The segregation driving force for 3$sp$ elements to the GB or open surface is the weakest for Al and increases from Al towards Mg or S. The preferred GB substitutional segregation site by Mg and Al is site 9, while Si, P, and S prefer segregation sites 1 or 3. 

The changes to $W_{sep}$ and $G_{disl}$ induced by the presence of different impurity atoms at the GB or USF modeled using supercells are listed in Table \ref{tbl:wkg-dil}. We analyze the computed changes via a procedure similar to that by Lozovoi \textit{et. al.} \cite{lozovoi2006structural,lozovoi2008boron} by defining several reference systems (relaxed or unrelaxed atomic configurations with or without an impurity atom in the bulk-like environment or in the defective region) to separate the total impurity effect into a chemical and a strain contribution (the procedure is detailed in the Supplemental Material). The results of this analysis are also presented in Table \ref{tbl:wkg-dil}. 

For 3$sp$ impurity elements, the calculated $\Delta W_{sep}$ is the highest for Al and decreases from Al towards Mg or S; the same trend is exhibited by the segregation energy to a GB or to an open surface. This correlation is natural since $W_{sep}$ is the difference between the surface energy (taken twice) and the GB energy. Table \ref{tbl:wkg-dil} also shows that the strain contribution to $\Delta W_{sep}$ is small compared to the chemical contribution, so that the impurity effect on the work of separation is predominantly determined by the chemical contribution. Messer and Briant \cite{Messmer1982457} suggested that more electronegative impurities should lower $W_{sep}$. This trend is partly confirmed by the present calculations: elements Mg and Al that are less electronegative than Cu increase the work of separation while more electronegative elements P and S decrease the $W_{sep}$.  However, the variation of $\Delta W_{sep}$ along the $3sp$ series does not quite follow the same trend as the electronegativity difference $\Delta \chi$ between the impurities and Cu. Thus, although Si is more electronegative than Cu, it is found to increase $W_{sep}$. One reason for the observed deviation from the expected behavior may be the existence of surface states due to the sudden termination of the crystal atomic order at either a surface or a GB. Rodriguez \textit{et al.} \cite{Rodriguez} have shown experimentally that the charge transfer (related to electronegativity difference) between impurities and host atoms at the surface is very different from that for the atoms in the bulk.  

Table \ref{tbl:wkg-dil} also shows that the strain contribution to $\Delta G_{disl}$ is very large and negative for 3$sp$ impurities in Cu. As the interatomic bonds are expected to be stronger in the bulk than in the USF region, the strain contribution to $\gamma_{usf}$ and $\Delta G_{disl}$ is expected to be negative. The chemical contribution is large and positive; it regularly decreases with the increasing electronegativity from Mg to S. When added together, the strain and the chemical contributions to $\Delta G_{disl}$ nearly balance each other, so that the overall impurity effect on the work of dislocation emission is calculated to be relatively small. 

As discussed above, the impurity-induced changes to both $W_{sep}$ and $G_{disl}$ are calculated to vary quite regularly along the $3sp$ series from Mg to S, so that the reason why P and S should have such contrasting effects on the ductility of Cu is not obvious \cite{henderson92tr,sandstrom2013influence}. However, if we plot $W_{sep}$ and $G_{disl}$ as a function of the segregation density $\Gamma$ in Fig. \ref{fig:energ-den}, a striking feature may be seen that only S can decrease the $W_{sep}$ below the level of $G_{disl}$. For Mg, Al, Si and P the value of $W_{sep}$ always stays above that of $G_{disl}$, no matter how high the segregation density is. At the same time, P segregation significantly decreases the value of $G_{disl}$. The resulting lowering of the activation barrier for dislocation emission from the tip of a GB crack is consistent with the experimentally observed positive effect of P alloying on the ductility of polycrystalline copper \cite{sandstrom2013influence,sandstromrole}. In Fig. \ref{fig:energ-den}, the extrapolation of the calculated results to high $\Gamma$ values is justified by the linearity of the obtained concentration dependencies, which suggests that the impurity--impurity interactions mostly cancel out in the $G_{disl}$ and $W_{sep}$ calculations. 

Whether or not a GB in Cu can become so enriched in S as to reverse the order of $W_{sep}$ and $G_{disl}$ is another question. The segregation energies listed in Table \ref{tbl:seg} indicate that 3$sp$ impurities prefer the $\Sigma 5$ grain boundary to the stacking faults. The segregation energy can vary with increasing the segregation density as a result of mutual interactions among the segregating impurities. Attractive interactions lower the impurity chemical potential at the GB and thus promote further impurity segregation, while repulsive interactions counteract the segregation to terminate further GB enrichment at the point of equilibrium where the repulsion is so strong that the incoming impurity atom will gain more energy elsewhere in the crystal (for instance, at SFs).

\begin{figure}[htb]
\centering
\includegraphics[width=0.5\textwidth]{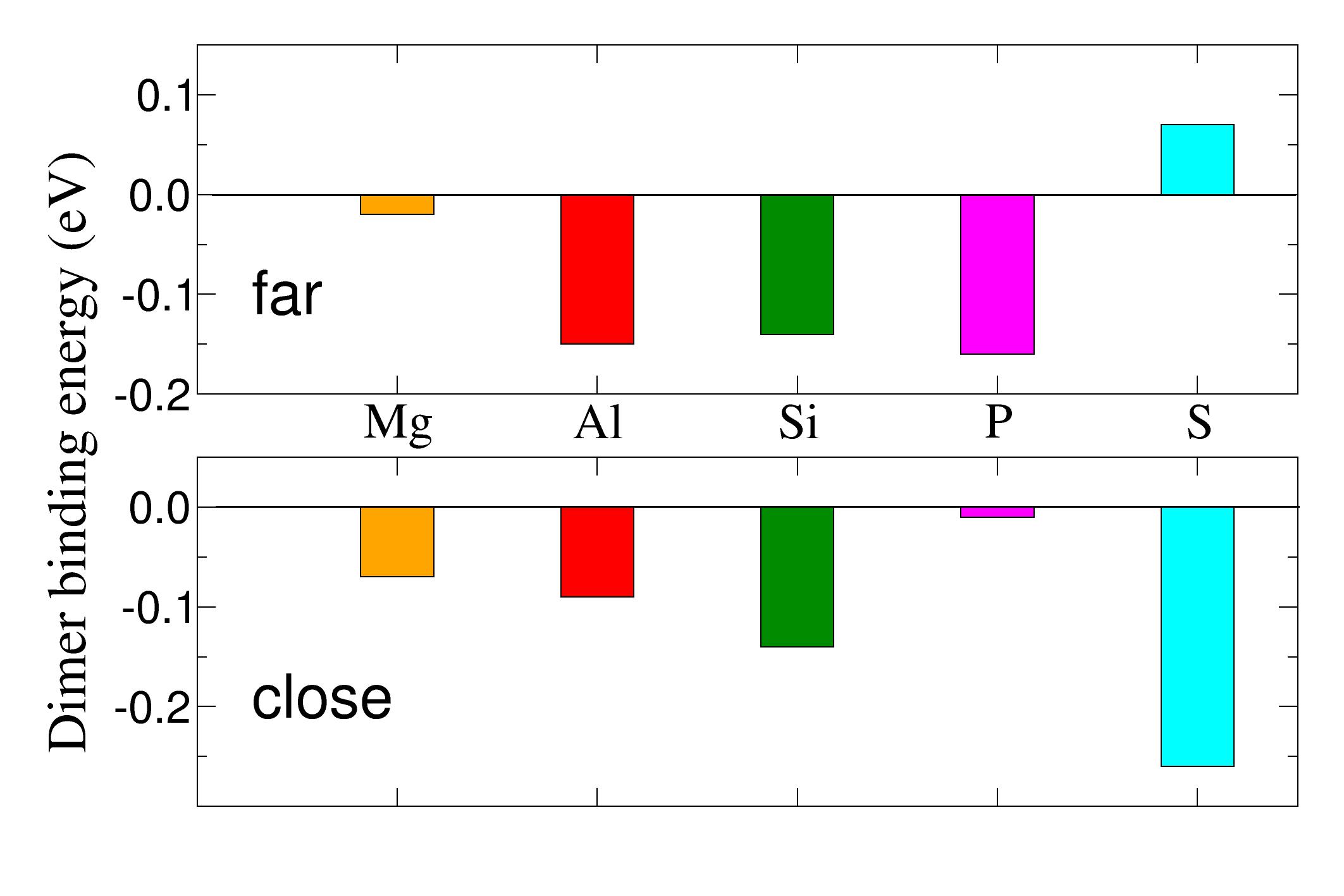}
\caption{The strongest binding energy for dimers at the $\Sigma 5$ GB of copper in which the impurities are separated by far (5.3--7.3 \AA) or close (2.4--3.1 \AA) distances. \label{fig:dimbind}}
\end{figure}
\begin{table}[htb]
\centering
 \caption{The impurity interaction energy (eV) of 3$sp$ impurities at GB for two different values of segregation density $\Gamma$.  \label{tbl:selfinter}}
\begin{tabular}{lccccc}
\hline
  \multirow{2}{*}{$\Gamma$, nm$^{-2}$} &  \multicolumn{5}{c}{Self-interaction energy (eV)} \\
\cline{2-6}
 &Mg &Al &Si &P &S \\
\hline  
1.196 & -0.20 & -0.24 & -0.2 & -0.22 & -0.19\\
2.393 & -0.26 & -0.11 & 0.00 & 0.02 & -0.10  \\
\hline   
\end{tabular}
\end{table}

To illustrate the interactions between impurity atoms, we calculated the total energy for dimers of impurities considered at different sites and at the $\Sigma 5$ GB thus searching for the most likely configurations of the dimers. The dimer binding energy was calculated by subtracting from the segregation energy of a dimer (Table SM3) the segregation energies of two single atoms (Table \ref{tbl:seg}) calculated using the same supercell by a straightforward formula $E_b=\Delta\varepsilon_{ij}^{dimer} - \Delta\varepsilon^{single}_i - \Delta\varepsilon^{single}_j$ (superscripts $i$ and $j$ indicate the sites occupied by impurities). We briefly summarize the calculated data in Fig. \ref{fig:dimbind} for the impurity pairs with the strongest binding at either far (5.3--7.3 \AA) or close (2.4--3.1 \AA) distances, respectively. The Figure clearly shows that a S dimer tends to be closely bound, while a P dimer tends to be dissociated.

However, the dimer binding energy cannot represent the total interaction among the impurity atoms forming a dense segregation because of its pairwise nature (many-body effects are neglected) and the limited range (by the size of GB supercell). To calculate an accurate interaction energy, we have to introduce another formalism to treat segregation, in which the impurity density in the bulk region and at the GB is the same. We refer to this treatment of segregation as the \textit{fixed-composition formalism} (here, \textit{fixed-composition} refers to the total impurity content in the system), while the previous one is hereafter called the \textit{variable-composition formalism}. In the \textit{fixed-composition formalism}, the segregation energy is computed as the energy difference between the GB structures with (the same amount of) impurities at GB sites as well as at bulk sites, and therefore it represents just the binding energy of the impurities to the GB, while the interactions among the impurities are mostly cancelled out. The segregation energies for one and two impurity atoms in the GB supercell for the \textit{fixed-composition case} are collected in Tables SM4 and SM5, respectively. 

By subtracting the segregation energy of the \textit{fixed-composition case} from that of the  \textit{variable-composition case}, we collect the impurity interaction energy in Table \ref{tbl:selfinter}. One can see that Mg and S show stronger attractive interactions compared to Al, Si and P impurities. The S--S dimer has an attractive interaction no matter what positions the two S atoms occupy at the GB. So, S tends to densely segregate and the segregation density can exceed the critical value where $W_{sep}$ falls below $G_{disl}$. At the same time, P shows a less attractive interaction than S and tends to distribute evenly. We performed a different calculation to independently confirm that the segregation of S can exceed the critical density, in which we placed 10 S atoms in the GB supercell (corresponding to $\Gamma$=11.96/nm$^2$) to obtain a segregated energy of -0.67 eV/atom. 

We therefore conclude that S can densely segregate to GBs and induce a ductile-to-brittle transition in GB deformation behavior. In contrast, P is predicted not to embrittle Cu, no matter how densely it segregates at GBs. On the other hand, P may compete with S for segregation sites at GBs, which can counteract the embrittlement caused by densely segregated S \cite{korzhavyi1999theoretical}.  

In summary, we investigated the nature of 3$sp$ impurity-mediated changes in the GB deformation behavior of polycrystalline Cu. The analyses suggest that $\Delta W_{sep}$ is related to the electronegativity  difference between Cu and the impurity. $\Delta G_{disl}$ has considerable contributions due to both the chemical and strain effects. The sharp contrast between the effects of P and S on the ductility on polycrystalline Cu is just a result of a delicate balance of the basic physical contributions. 

This work was funded by the Swedish Nuclear Fuel and Waste Management Company (SKB) and the Swedish Foundation for Strategic Research (SSF, project ALUX). The research used resources provided by the Swedish National Infrastructure for Computing (SNIC) at the National Supercomputer Center (NSC), Link\"oping, the High Performance Computing Center North (HPC2N), and at the PDC Center for High-performance Computing, Stockholm.

\bibliography{thesis}

\end{document}